\documentclass[prl,twocolumn,superscriptaddress,amsmath,amssymb,floatfix,noshowpacs]{revtex4-2} 
\usepackage{graphicx}   
\usepackage{dcolumn}    
\usepackage{bm}         
\usepackage{hyperref}   
\usepackage[toc,page]{appendix}

\begin{document}

\title{\textbf{Quantum interfaces with multilayered superwavelength atomic arrays}}
\author{Roni Ben-Maimon}
\affiliation{Department of Chemical \& Biological Physics, Weizmann Institute of Science, Rehovot 7610001, Israel}
\author{Yakov Solomons}
\affiliation{Department of Chemical \& Biological Physics, Weizmann Institute of Science, Rehovot 7610001, Israel}
\author{Nir Davidson}
\affiliation{Department of Physics of Complex Systems, Weizmann Institute of Science, Rehovot 7610001, Israel}
\author{Ofer Firstenberg}
\affiliation{Department of Physics of Complex Systems, Weizmann Institute of Science, Rehovot 7610001, Israel}
\author{Ephraim Shahmoon}
\affiliation{Department of Chemical \& Biological Physics, Weizmann Institute of Science, Rehovot 7610001, Israel}
\date{\today}

\begin{abstract}
We consider quantum light-matter interfaces comprised of multiple layers of two-dimensional atomic arrays, whose
lattice spacings exceed the wavelength of light. While the coupling of light to a single layer of such a ‘superwavelength’ lattice is considerably reduced due to scattering losses to high diffraction orders, we show that the addition of layers can suppress these losses through destructive interference between the layers. Mapping the problem to a 1D model of a quantum interface wherein the coupling efficiency is characterized by a reflectivity, we analyze the latter by developing a geometrical optics formulation, accounting for realistic finite-size arrays. We find that optimized efficiency favors small diffraction-order angles and small interlayer separations, and that the coupling inefficiency of two layers universally scales as $N^{-1}$ with the atom number per layer $N$. We validate our predictions using direct numerical calculations of the scattering reflectivity and the performance of a quantum memory protocol, demonstrating high atom-photon coupling efficiency. We discuss the utility of our technique for applications in tweezer atomic arrays platforms.
\end{abstract}

\maketitle

The ability to establish an efficient interface between photons and atoms is of fundamental importance in quantum science and is at the basis of a variety of quantum applications, from quantum memories and networks to entanglement generation and photonic many-body physics \cite{ref55,ref15,ref17,ref19,ref64}
A quantum interface is aimed to efficiently couple an atomic system to a specific target photon mode, in which light is sent and collected. The corresponding coupling rate $\Gamma$ should then exceed the loss rate $\gamma_{\mathrm{loss}}$ due to scattering to other modes. Accordingly, the efficiency of the quantum interface in performing various quantum tasks can be characterized by the ratio $r_0=\Gamma/(\Gamma+\gamma_{\mathrm{loss}})$ \cite{Uni}.

\begin{figure}[t]
  \centering
  \includegraphics[width=\columnwidth]{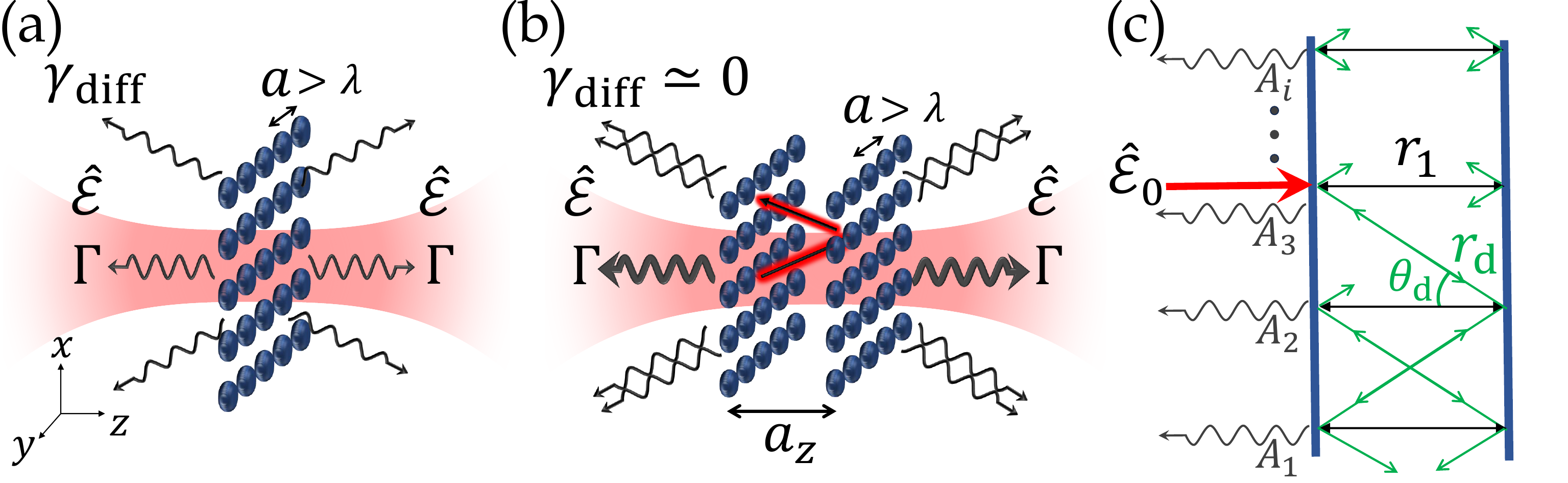}
  \caption{(a) A superwavelength 2D atomic array is coupled to a normal-incident target mode $\hat{\mathcal{E}}$ at rate $\Gamma$ while exhibiting scattering losses at rate $\gamma_{\text{diff}}$ to higher diffraction orders. (b) Adding more layers of superwavelength 2D arrays can suppress the losses $\gamma_{\text{diff}}$ due to destructive interference of the higher diffraction orders. (c) Geometrical optics theory of multiple ray scattering between array layers. Rays scatter to the normal zeroth diffraction order with amplitude $r_1$ and to higher diffraction orders with amplitude $r_{\mathrm{d}}$ at an angle $\theta_{\mathrm{d}}$. For a layer of finite linear size $L$, there are $M\sim(L/a_z\tan\theta_{\mathrm{d}})^2$ relevant points at which rays arrive and scatter, whose corresponding field amplitudes $A_i$ ($i\in{1,...,M}$) can be calculated semi-analytically.}\label{Fig1}
\end{figure}

In typical platforms, comprising single atoms or dilute ensembles, the losses $\gamma_{\mathrm{loss}}$ are dominated by unavoidable individual-atom spontaneous scattering. Hence, high efficiency $r_0$ typically requires to enhance the coupling $\Gamma$ to the target mode,\emph{ e.g.}, by placing the atomic system in a cavity or waveguide \cite{ref57,ref58,ref59,ref61}, or by using elongated atomic ensembles \cite{ref51,ref53,ref54}. An alternative approach would be to inhibit the scattering losses $\gamma_{\mathrm{loss}}$, which can be achieved in free space by using two-dimensional (2D) spatially ordered atomic arrays. In a subwavelength array, where the array lattice spacing $a$ is smaller than the relevant optical wavelength $\lambda$, the combination of spatial order and collective response of array atoms leads to a directional coupling to the target mode and hence to drastically reduced losses $\gamma_{\mathrm{loss}}\ll \Gamma$ \cite{ref28,ref27,Efi2017,ref29,ref30,ref31,ref32,ref37,ref38,ref40,ref18,ref41,ref42,ref43,ref44,ref45,ref46,ref47,ref26,ref48}. However, for ``superwavelength" arrays, where $a>\lambda$, radiative diffraction orders emerge \cite{Uni,Efi2017}, leading to significant losses and reduction of the efficiency $r_0$ (Fig. \ref{Fig1}a, with $\gamma_{\mathrm{loss}}=\gamma_{\mathrm{diff}}$). This poses a particulary crucial problem for quantum light-matter applications based on tweezer atomic arrays, which are typically superwavelength \cite{ref02,ref04,ref05,ref06,ref08,ref09,ref11,ref14,ref65}.

Here we show how efficient coupling can be achieved also in the superwavelength case by considering multiple array layers: destructive interference of radiative diffraction orders emanating from different layers leads to their cancellation (Fig. 1b). Mapping the problem onto a 1D scattering model allows us to link the efficiency $r_0$ to the array reflectivity, thus optimizing the latter by designing configurations of destructively interfering diffraction orders. Finite-size effects are analyzed by developing a geometrical optics theory, corroborated by numerical scattering and quantum-memory simulations. We find a universal scaling of the efficiency with the array size, and reveal that high efficiencies require small diffraction-order angles and small interlayer separations. We consider examples of square and triangular lattices, beginning with the cancellation of a single existing set of radiative diffraction orders by using two layers, obtaining efficiencies $r_0>0.99$ with thousands of atoms. For two existing sets of diffraction orders, we discuss their approximate or exact cancellation using two or four layers, respectively, guided by an intuitive geometrical approach.

\emph{The system.---} We consider an atomic array of identical two-level atoms, comprised of $N_{z}$ identical layers spread along the $z$-axis with interlayer spacing $a_z$.  Each layer $l\in\left\{ 0,1,\dots N_{z}-1\right\}$ forms a 2D lattice on the $xy$ plane, with a lattice spacing $a$ and an angle $\psi$ between the primitive translation vectors ($\psi=\pi/2,\pi/3$ for square and triangular lattices, respectively). Considering a possible lateral shift in the origin of each layer, $\mathbf{b}_{l}=\left(x_{l},y_{l},0\right)$, the atomic positions are
$\mathbf{r}_{\mathbf{n},l}=\left(n_1 a+n_2a\cos\psi,n_2 a\sin\psi ,l a_{z}\right)+\mathbf{b}_{l}$, with $n_{1},n_{2}\in\{1,\dots\sqrt{N}\}$.
The array can be illuminated from either side by a weak normal-incident beam, whose wavelength $\lambda=2\pi/k$ is assumed smaller than the intralayer lattice spacing, $a>\lambda$, making the array superwavelength (Fig. 1).

Writing the Heisenberg-Langevin equations for the atomic lowering operators $\hat{\sigma}_{\mathbf{n}l}$ within a Born-Markov approximation, we obtain linearly responding atoms coupled by photon-mediated dipole-dipole interactions \cite{ref26,ref48}. These equations can be cast in the form of an effective interaction between collective dipoles $\hat{P}_{l}=\frac{1}{\sqrt{N}}\sum_{\mathbf{n}}\hat{\sigma}_{\mathbf{n}l}$ of layers $l$ \cite{Uni},
\begin{eqnarray}
&&\dot{\hat{P}}_{l}=\ i\delta\hat{P}_{l} -\sum_{l=0}^{N_{z}-1}D_{ll'}\hat{P}_{l'}+i\sqrt{\Gamma_0}\hat{{\cal E}}_{0,l}+\hat{F}_{\text{diff},l},
\nonumber\\
&&\hat{{\cal E}}_{\pm}(z)=\hat{{\cal E}}_{0,\pm}(z)+i\sqrt{\frac{\Gamma_{0}}{2}}e^{\pm ikz}\sum_{l=0}^{N_{z}-1}e^{\mp ikla_{z}}\hat{P}_{l}.
\label{eq_EOM_layers}
\end{eqnarray}
Here we assumed each layer is represented by an infinite 2D lattice. Correspondingly, the mode of the normal-incident field, described by the operator $\hat{{\cal E}}_{\pm}$ and its respective input $\hat{{\cal E}}_{0,\pm}$ ($\pm$ denoting right or left propagation) is taken to be a plane wave. Finite-size effects are treated further below. We observe that the collective dipole $\hat{P}_l$ is directly coupled to the normal-incident field from both sides $\hat{{\cal E}}_{0,l}=[\hat{{\cal E}}_{0,+}(a_{z}l)+\hat{{\cal E}}_{0,-}(a_{z}l)]/\sqrt{2}$ at a rate $\Gamma_0=\frac{3}{4\pi}\frac{\lambda^2}{a^2\sin\psi}\gamma$, with $\gamma$ the individual-atom spontaneous emission rate. Its coupling to the collective dipoles of others layers $l'$ is given by the interaction kernel,
\begin{eqnarray}\label{eq_DD_kernel}
        \displaystyle
D_{ll'}=\sum_{\mathbf{m}}\frac{\Gamma_{\mathbf{m}}}{2}e^{ik_{z}^{\mathbf{m}}a_{z}\left|l-l'\right|}e^{-i\mathbf{Q}_{\mathbf{m}}\cdot\left(\mathbf{b}_{l}-\mathbf{b}_{l'}\right)},
\end{eqnarray}
mediated by all diffraction orders $\mathbf{m}=\left(m_{1},m_{2}\right)$ ($m_{1},m_{2}\in\mathbb{Z}$), with reciprocal lattice momentum $\mathbf{Q}_{\mathbf{m}}=\frac{2\pi}{a}\left(m_{1},-m_1 \cot\psi+m_2\frac{1}{\sin\psi}\right)$, longitudinal wavevector  $k_{z}^{\mathbf{m}}=k\sqrt{1-|\mathbf{Q}_{\mathbf{m}}|^{2}/k^2}$, and coupling rates $\Gamma_{\mathbf{m}}=\Gamma_{0}\frac{1-|\mathbf{Q}_{\mathbf{m}}\cdot\mathbf{e}_{\mu}|^{2}/k^2}{k_{z}^{\mathbf{m}}/k}$, where $\mathbf{e}_{\mu}$ is the orientation of the atomic dipole (taken circularly polarized). For a single layer, the radiation rate $2\mathrm{Re}[D_{ll}]$ then includes the coupling $\Gamma_0$ to the normal-incident mode of interest $\mathbf{m}=0$, accompanied by scattering losses $\Gamma_{\mathbf{m}}$ and corresponding vacuum noise $\hat{F}_{\text{diff},l}$ due to higher radiative diffraction orders $|\mathbf{m}|>0$ satisfying $|\mathbf{Q}_{\mathbf{m}}|<k$. The losses $\Gamma_{\mathbf{m}}\sim\mathcal{O}(\Gamma_0)$ thus preclude a single superwavelength layer from exhibiting high coupling efficiencies \cite{YakovCavity}.

\emph{1D model of quantum interface.---}
Examining  Eq. (\ref{eq_EOM_layers}), it becomes evident that the fields ${\cal \hat{E}}_{\pm}\left(z\right)$ are coupled to multilayer collective dipoles $\hat{P}_{\pm}=\frac{1}{\sqrt{N_z}}\sum_{l=0}^{N_{z}-1}e^{\mp ikla_{z}}\hat{P}_{l}$. Assuming that the interlayer spacing satisfies
$a_{z}/\lambda=\mathbb{N}/2$, we obtain a single collective dipole $\hat{P}=\hat{P}_+=\hat{P}_-$ coupled symmetrically to the field on both sides. A full description of Eqs. (\ref{eq_EOM_layers}) in terms of this symmetric dipole mode $\hat{P}$ requires that it is an eigenmode of the interaction kernel $D_{ll'}$. We begin with a two-layer system, where this requirement is always satisfied, and discuss multiple layers further below. We obtain,
\begin{eqnarray}
&&\dot{\hat{P}}=\left[i\left(\delta-\Delta\right)-\frac{\Gamma+\gamma_{\text{diff}}}{2}\right]\hat{P}+i\sqrt{\Gamma}{\cal \hat{E}}_{0}\left(0\right)+\hat{F}_{\text{diff}},
\nonumber\\
&&\hat{{\cal E}}\left(z\right)=\hat{{\cal E}}_{0}\left(z\right)+i\sqrt{\Gamma}\hat{P},
\label{eq_EOM_col}
\end{eqnarray}
with $\Delta$ given by the collective shift of a single layer \cite{Efi2017,sup}. These equations exhibit the form of a generic 1D model of an atom-photon interface \cite{Uni}: the collective dipole $\hat{P}$ is coupled at rate $\Gamma$ to the symmetric superposition $\hat{{\cal E}}=[\hat{{\cal E}}_+ +\hat{{\cal E}}_-]/\sqrt{2}$ of the normal-incident ``target mode", at which light can be sent and collected from both sides. In addition, losses due to scattering to higher radiative diffraction orders exist at rate $\gamma_{\text{diff}}$ and accompanied by vacuum noise $\hat{F}_{\text{diff}}$. For two layers we obtain
\begin{eqnarray}\label{eq_C_params}
        \displaystyle
\Gamma=2\Gamma_{0},\ \gamma_{\text{diff}}=\sum_{\underset{\quad\neq00}{ \mathbf{m} \in R,}}\Gamma_{\mathbf{m}}\left(1+e^{ika_{z}}e^{ik_{z}^{\mathbf{m}}a_{z}}e^{i\mathbf{Q}_{\mathbf{m}}\cdot\mathbf{b}_{1}}\right),
\end{eqnarray}
where the domain $R$ in the sum includes only the radiating diffraction orders, \emph{i.e.}  $\mathbf{m}$ satisfying $\left|\mathbf{Q}_{\mathbf{m}}\right|<k$. Within the 1D model description (\ref{eq_EOM_col}), the coupling efficiency of the quantum interface, and hence of various quantum tasks such as memory and entanglement generation, is fully characterized by the quantity \cite{Uni}
\begin{eqnarray}\label{eq_r0}
        \displaystyle
r_{0}=\frac{\Gamma}{\Gamma+\gamma_{\text{diff}}}.
\end{eqnarray}
Importantly, the efficiency $r_0$ is equal to the on-resonance reflectivity of the target-mode light sent from one direction and reflected back by the array \cite{Uni}. We will therefore concentrate on optimizing the array's reflectivity.

\emph{Resonant spacings.---} To optimize $r_0$, it is imperative to minimize $\gamma_{\text{diff}}$. This can be achieved by choosing the phases in Eq. (\ref{eq_C_params}) to ensure destructive interference between the layers, yielding a set of conditions on array spacings and lateral shift $\left(a_{z},a,\mathbf{b}_{1}\right)$. We begin by examining scenarios where only the first diffraction order is radiative and then extend the discussion to higher orders.
\begin{figure}[t]
  \centering
  \includegraphics[width=\columnwidth]{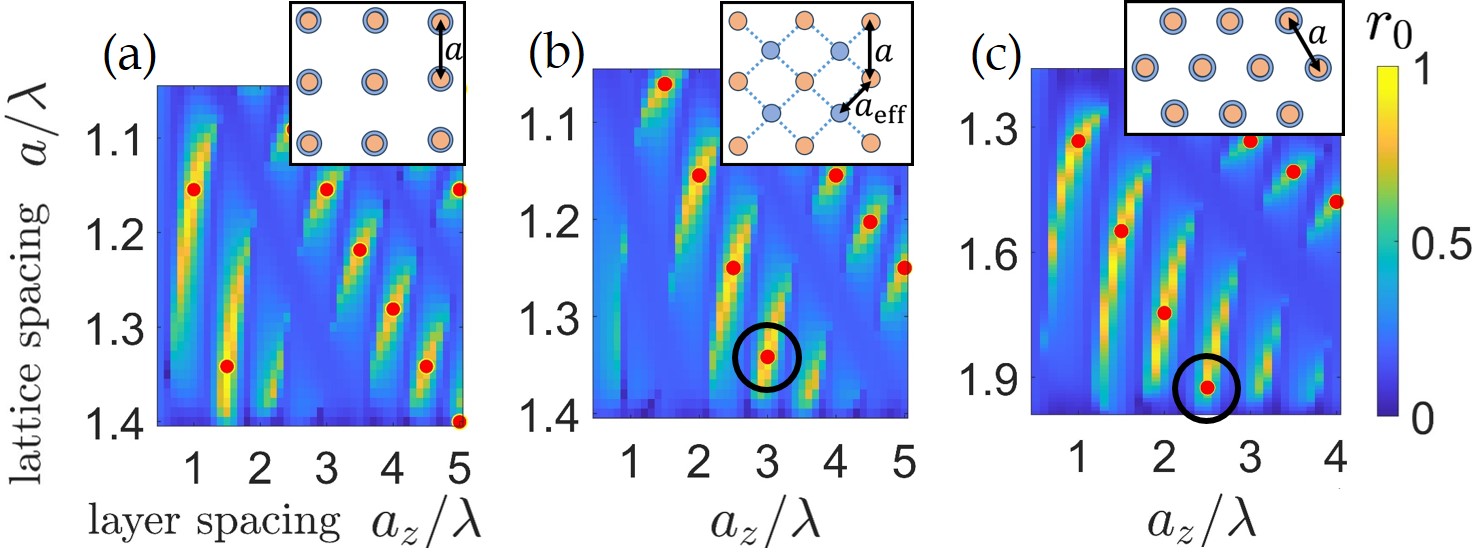}
  \caption{
  Reflectivity map of two-layer arrays obtained numerically as a function of the lattice and interlayer spacings $a$ and $a_z$.
  Reflectivity maxima agree with the analytically predicted resonant-spacing configurations (red dots). (a) Square lattice, non-shifted configuration $\mathbf{b}_1=0$ (for $N=35\times35$ atoms per layer and incident beam waist $w=14\lambda$). (b) Same as (a) for shifted configuration $\mathbf{b}_{1}=\left(\frac{a}{2},\frac{a}{2}\right)$. (c) Triangular lattice (no shift $\mathbf{b}_1=0$, for $N=492$ and $w=8\lambda$).}
  \label{Fig2}
\end{figure}
Consider first a superwavelength square lattice satisfying $a<\sqrt{2}\lambda$, where the first diffraction order consists of 4 diffraction modes $\mathbf{m}=(\pm1,0),(0,\pm1)$. In the non-shifted configuration $\mathbf{b}_{1}=0$, $\gamma_{\text{diff}}=0$ is achieved by ensuring that the longitudinal phases of the target mode and the radiating diffraction order accumulate opposing signs, $e^{ika_{z}}e^{ik_{z}^{\mathbf{m}}a_{z}} = -1$, resulting in destructive interference among the diffraction modes. Satisfying this condition on the longitudinal phases yields the sets of ``resonant spacings" $\left(a_{z},a\right)$ plotted in Fig. \ref{Fig2}a.
The predicted sets are indeed seen to coincide with maxima of the reflectivity obtained from a direct numerical simulation \cite{Efi2017} of the scattering off the array of a right-propagating incident Gaussian beam at resonance $\delta=\Delta$ (Fig. \ref{Fig2}a).
In contrast, in the shifted configuration, where the chosen lateral shift $\mathbf{b}_{1}=\left(\frac{a}{2},\frac{a}{2}\right)$ yields a $\pi$-phase shift between the layers, destructive interference requires matched longitudinal phases, $e^{ika_{z}}e^{ik_{z}^{\mathbf{m}}a_{z}} = 1$ (Fig. \ref{Fig2}b). Such an arrangement lends itself to a geometrical interpretation wherein both layers create a denser array on a single effective plane, with an effective subwavelength lattice spacing $a_{\mathrm{eff}}=a/\sqrt{2}<\lambda$ (inset). This intuitive picture serves as a useful guide for scaling up our approach to multiple diffraction orders and layers (see below). 

\emph{Geometrical optics theory.---} We now turn to develop a geometrical optics formulation of the multiple scattering of rays between the array layers, useful for analyzing finite-size arrays (Fig. \ref{Fig1}c). To this end, we first use the full wave theory \cite{Efi2017} to find the scattering amplitudes off a single layer to the normal-incident and higher diffraction orders, given by $r_{1}=-\Gamma_{0}/\left(\Gamma_{0}+\gamma_{\text{diff}}^{1}\right)$ and $r_{\mathrm{d}}=-\gamma_{\text{diff}}^{1}/\left(\Gamma_{0}+\gamma_{\text{diff}}^{1}\right)$, respectively, with  $\gamma_{\text{diff}}^{1}=\sum_{\mathbf{m}\in R,\neq00}\Gamma_{\mathbf{m}}$ \cite{sup}. Then, taking a geometrical optics approximation, we neglect additional wave-diffraction effects by summing over all ray trajectories to obtain the reflectivity $r_0$. For two infinite layers we perform the infinite sum analytically, recovering the reflectivity maps from Fig. \ref{Fig2} \cite{sup}.
For layers of finite linear size $L$ ($L=a\sqrt{N}$ in a square lattice) and given the angle $\theta_{\mathrm{d}}=\arcsin(|\mathbf{Q}_{\mathbf{m}}|/k)$ of the first-emerging diffraction order, there are $M\sim(L/a_{z}\tan\theta_{\mathrm{d}})^2$ points $(x_i,y_i)$ ($i=1,...,M$) on each layer at which rays arrive and re-scatter (Fig. 1c).
Field amplitudes $A_i$ at these points evolve between round trips via a single-layer ray-scattering matrix $\mathcal{S}_{\mathrm{r}}$ built from $r_1$ and $r_{\mathrm{d}}$, yielding $\vec{A}=\mathcal{S} \vec{A}_{\mathrm{in}}$ with
\begin{eqnarray}
\mathcal{S}=r_1\left[1+\mathcal{S}_{\mathrm{t}}\left(1-\mathcal{S}_{\mathrm{r}}^{2}\right)^{-1}\mathcal{S}_{\mathrm{t}}\right], \quad \mathcal{S}_{\mathrm{t}}=\mathcal{S}_{\mathrm{r}}+e^{ik a_z},
\label{S}
\end{eqnarray}
and $A_{\mathrm{in},i}=\delta_{i,(M+1)/2}$ \cite{sup}.
Finally, the reflectivity is given by $r_0=\sum_i A_i e^{-(x_i^2+y_i^2)/(2w^2)}$, with $w$ being the incident beam waist, and can be estimated semi-analytically \cite{sup}. This further allows us to analytically find a universal asymptotic scaling $N^{-1}$ of the coupling inefficiency $1-r_0$ in two-layer arrays with the atom number per-layer $N$ \cite{sup}, as demonstrated in Fig.~3.

\begin{figure}[t!]
  \centering
  \includegraphics[width=\columnwidth]{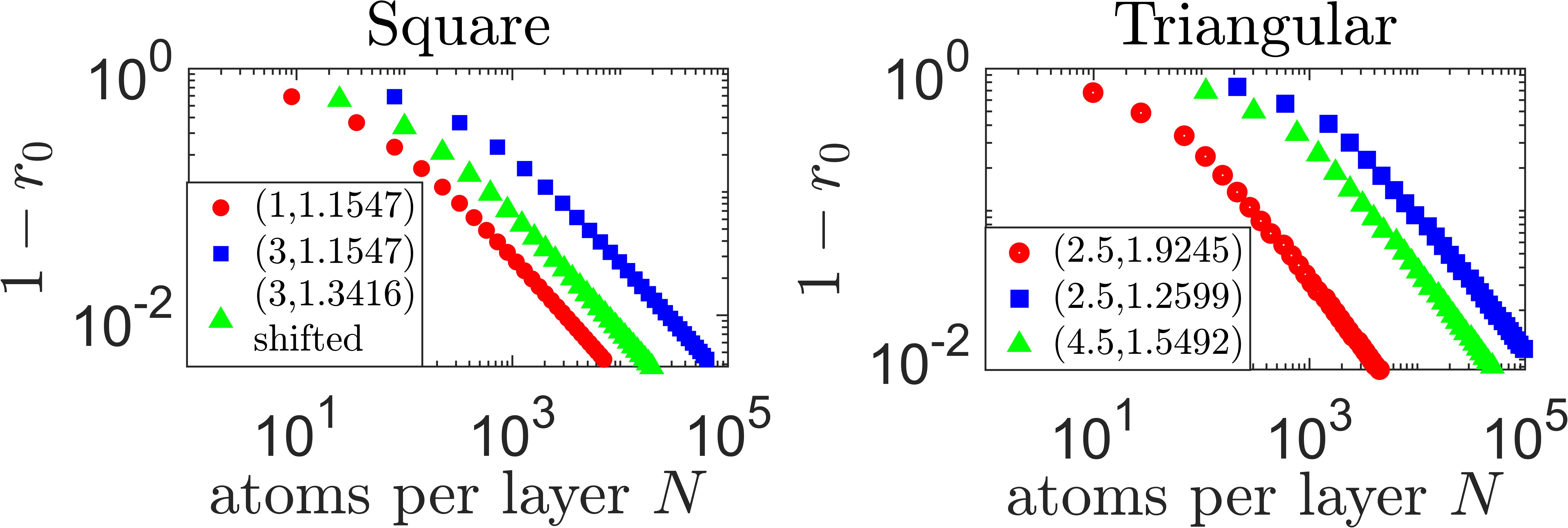}
  \caption{Coupling inefficiency $1-r_0$ of two-layer arrays calculated semi-analytically by the geometrical optics theory. Various optimal configurations $\left(a_z,a\right)/\lambda$ from Fig. 2 all exhibit the universal asymptotic scaling $N^{-1}$ with the atom number per layer $N$ (reaching $N^{-0.98}$ for displayed $N$ values \cite{sup}).}\label{Fig3}
\end{figure}

\begin{figure}[t!]
  \centering
  \includegraphics[width=\columnwidth]{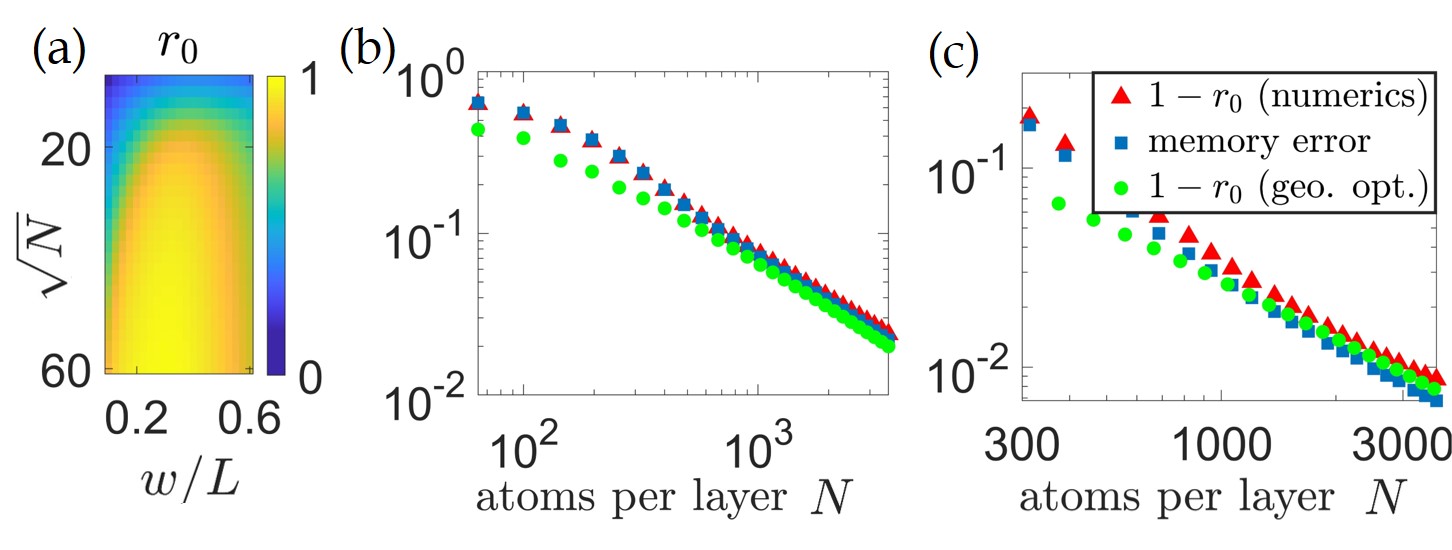}
  \caption{Coupling efficiency $r_0$ of a quantum interface formed by two-layer arrays. (a) Square array: $r_0$ obtained from numerical scattering calculations as a function of $\sqrt{N}$ and the beam waist $w/L$, with $N$ the number of atoms per layer and $L=\sqrt{N}a$ the layer length [shifted configuration $\left(a_z/\lambda=3,a/\lambda=1.3416\right)$ depicted in Fig. \ref{Fig2}b]. (b) Square array from (a): coupling inefficiency $1-r_{0}$ vs. $N$ for constant $w/L=0.3$, obtained in three independent ways:  numerical scattering, quantum memory infidelity, and geometrical optics theory. Power law fits (for $N\geq 576$) exhibit the scalings $N^{-0.91}$ (scattering) and $N^{-0.94}$  (memory), tending to the universal asymptotic scaling $N^{-1}$ predicted by the geometrical optics theory. (c) Triangular lattice: $1-r_{0}$ vs. $N$ for constant $w/L=0.3$ calculated as in (b), exhibiting the scalings (fitted for $N\geq 800$) $N^{-1.1}$ (scattering) and $N^{-1.08}$  (memory), again tending to the asymptotic prediction $N^{-1}$.  The observed inefficiencies are lower than those found for square arrays in (b) thanks to a lower diffraction angle $\theta_{\mathrm{d}}$ of the triangular lattice [configuration $\left(a_{z}/\lambda=2.5,a/\lambda=1.925\right)$ depicted in Fig. \ref{Fig2}c].
  }.\label{Fig4}
\end{figure}

\emph{Realistic quantum interfaces.---}
As a direct verification of the above predictions, we employ a numerical approach for the full quantum optical problem, including both wave-diffraction and realistic finite-sizes of arrays and target-mode Gaussian beams.
We begin by focusing on the specific set of optimal resonant spacings marked on Fig. \ref{Fig2}b. In Fig. 4a, we present the resulting coupling efficiency $r_0$ as a function of the beam waist $w$ and the layer size $N$, obtained by the direct numerical scattering calculation of the reflectivity described above. While $r_0$ increases with $N$, it exhibits an optimum waist $w/L$ within the range of $\left[0.3,0.4\right]$.
This optimum results from two competing effects: the reduction of the destructive interference between diffraction orders due to wave-diffraction of the finite beam is avoided by favoring large waists, while leakage of the beam at the boundaries of the finite layer is reduced for small waists.
The increase of $r_0$ with $N$ is studied in Fig. \ref{Fig4}b for a fixed waist $w/L=0.3$, revealing that the coupling inefficiency $1-r_{0}$ tends to the asymptotic scaling $N^{-1}$, as predicted by our geometrical optics theory.

We further demonstrate that $r_0$ indeed represents the efficiency of quantum tasks by performing a direct numerical calculation of the fidelity of a quantum memory protocol \cite{ref29}, applied to the same array and Gaussian target mode and detected symmetrically from both sides. The resulting memory infidelity indeed shows excellent agreement with $1-r_0$ (Fig. \ref{Fig4}b). Importantly, the observed agreement of the numerical results for both scattering and memory with the geometrical optics theory highlights that losses are dominated by the escape of rays out of finite-size layer boundaries, and not by diffraction effects that are neglected in geometrical optics.

The above conclusion implies that the efficiency $r_0$ increases with the number of ray round trips. The latter grows with the number of points to which rays arrive $M\propto(a_z \tan\theta_{\mathrm{d}})^{-2}$, favoring smaller interlayer spacing $a_z$ and diffraction angles $\theta_{\mathrm{d}}=\arcsin(|\mathbf{Q}_{\mathbf{m}}|/k)$. This motivates to consider a triangular lattice over the square one: while in a square lattice the second set of radiative diffraction orders appears at $a=\sqrt{2}\lambda$ ($\theta_{\mathrm{d}}=45^{\circ}$), in a triangular lattice it appears at $a=2\lambda$ ($\theta_{\mathrm{d}}\approx 35^{\circ}$). Therefore, the regime of a single lossy diffraction order extend to larger $a$ and hence to smaller $\theta_{\mathrm{d}}$ in a triangular lattice.
Considering two layers of a triangular lattice in this regime, we theoretically predict as before the sets of resonant spacings $\left(a_{z},a\right)$, which are seen to agree with the numerical scattering calculation (Fig. \ref{Fig2}c). Choosing a specific set with the largest $a$ and hence smallest angle $\theta_{\mathrm{d}}$, we plot in Fig. \ref{Fig4}c, for a constant waist $w/L=0.3$, the direct numerical results of the coupling inefficiency $1-r_0$ and the quantum memory error as a function of $N$, noticing their agreement. The inefficiency again exhibits the asymptotic decrease with $N$ predicted by the geometrical optics theory, however with lower values than those of the square array in Fig. 4b. This validates the superior performance of the triangular array as anticipated from the design principle revealed above. 

\begin{figure}[t!]
\centering
  \includegraphics[width=\columnwidth]{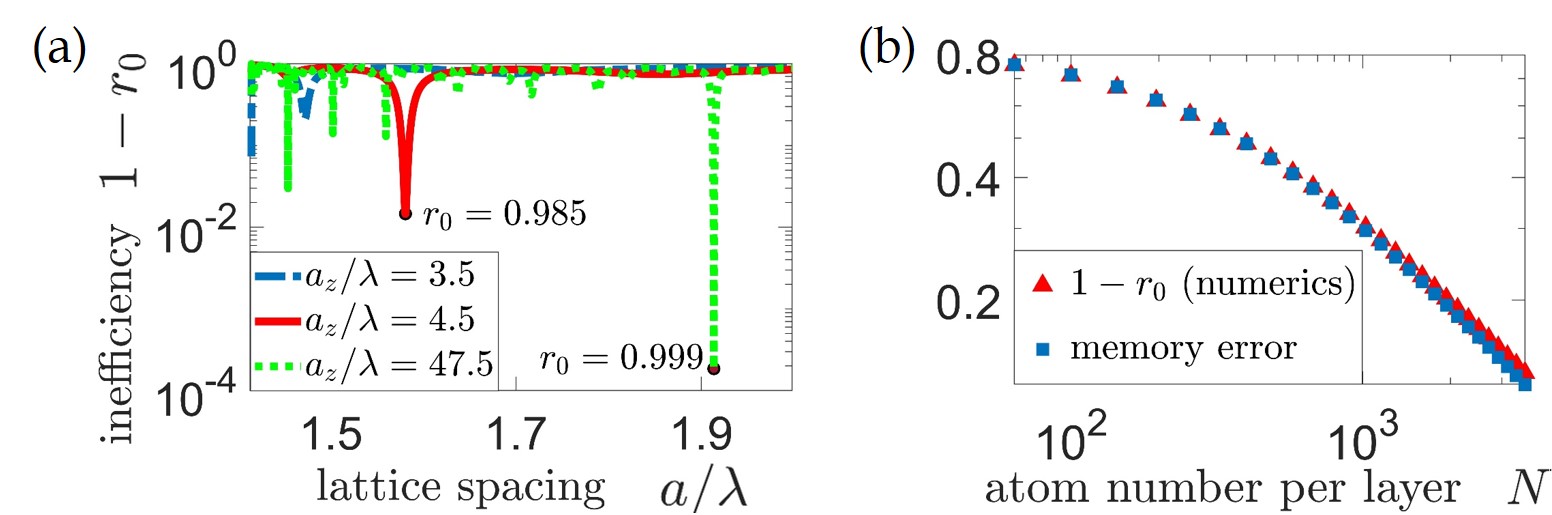}
  \caption{Approximate cancelation of two sets of diffraction orders with two layers. (a) Coupling inefficiency $1-r_0$ vs. lattice spacing $a$, for a shifted two-layer square lattice in the range $\sqrt{2}<a/\lambda<2$ wherein two sets of radiative diffraction orders exist ($|\mathbf{m}|=1,\sqrt{2}$). $r_0$ is calculated analytically from Eqs. (\ref{eq_C_params}), (\ref{eq_r0}) for varying interlayer spacings $a_z$. (b) Finite-array realization of the optimal configuration $\left(a_{z}/\lambda=4.5,a/\lambda=1.58\right)$: coupling inefficiency obtained numerically via scattering calculations and quantum memory infidelity.}\label{Fig5}
\end{figure}

\emph{Multiple diffraction orders and layers.---}
Larger lattice spacings $a$ introduce more sets of radiative diffraction orders beyond the first order discussed thus far.
Using only two layers for a simultaneous exact cancelation of these multiple diffraction orders becomes impractical due to growing number of constraints. Nevertheless, we can find realistic values of $(a_z,a)$ for which the destructive interference of multiple orders is approximately satisfied: that is, $\gamma_{\text{diff}}$ does not completely vanish but becomes small enough to yield high efficiencies $r_0\rightarrow 1$. This is illustrated in Fig. \ref{Fig5} for a two-layer square lattice with $\sqrt{2}<a/\lambda<2$ (containing two sets of radiative diffraction orders $|\mathbf{m}|=1,\sqrt{2}$). Figure \ref{Fig5}a shows $r_0$ obtained theoretically from Eq. (\ref{eq_r0}) for various $a_{z}$,
revealing enhanced reflectivity for $(a_z,a)$ values that approximately meet the conditions $k_{z}^{\mathbf{m}}a_{z}=2\pi\mathbb{N}$. As $a_{z}$ increases, more accurate theoretical solutions appear. However, recalling the design principle favoring smaller $a_z$, we choose a solution with moderate $a_z$ and employ the numerical scattering calculation to study the resulting reflectivity in a realistic, finite-size array. This is shown in Fig. \ref{Fig5}b, displaying the decrease of the inefficiency with the layer size $N$ which is nevertheless slower than that of the single diffraction order case (Fig. 4b).
This is mostly attributed to the larger diffraction angles associated with the second diffraction order, again demonstrating the design principle favoring smaller angles $\theta_{\mathrm{d}}$.

Exact cancellation of multiple diffraction orders can in principle be achieved, by adding more layers to the array. To this end, we are guided by the intuitive geometrical picture of Fig. 2b (inset), where lateral shifts $\mathbf{b}_l$ are used as additional degrees of freedom. We consider the case $e^{ika_{z}}e^{ik_{z}^{\mathbf{m}}a_{z}}=1$, wherein all the radiative diffraction orders $\mathbf{m}$ arrive at all layers with matched longitudinal phases, such that all layers are effectively placed on a single plane. Then, by properly choosing their lateral shifts, the layers appear together as an effective single subwavelength lattice, yielding $\gamma_{\text{diff}}=0$. For example, four layers are in principle able to eliminate $\gamma_{\text{diff}}$ for a square array with $\sqrt{2}<a/\lambda<2$ and for a triangular lattice with $2<a/\lambda<4/\sqrt{3}$. This requires finding sets of resonant pairs $(a_z,a)/\lambda$ that satisfy $e^{ika_{z}}e^{ik_{z}^{\mathbf{m}}a_{z}}=1$ for all radiative orders $\mathbf{m}$. We demonstrate this in \cite{sup} for the aforementioned square and triangular arrays, finding \emph{e.g.} $(8.5,1.55)$ and $(6.5,2.17)$, respectively.


\emph{Discussion.---} We have presented a detailed study of a quantum interface realized by multiple layers of a superwavelength array, establishing several design principles based on a geometrical optics perspective. This work reveals a promising approach for efficient coupling of light to the common platforms of tweezer atomic arrays, in applications such as fast quantum-state readout, quantum interconnection of atomic nodes, and quantum nonlinear optics. Multilayer tweezer arrays may be generated by the method from \cite{ref06}, while disorder imperfections can be analyzed as discussed previously \cite{Efi2017}. Going forward, this work motivates the extension of our geometric optics theory and the universal approach of symmetric array interfaces \cite{Uni}, to systematically account for multiple layers $N_z>2$ and non-symmetric coupling, exploring their potential benefits \cite{Darrick}.
\begin{acknowledgments}
We acknowledge fruitful discussions with Darrick Chang, and financial support from the Israel Science Foundation (ISF), the Directorate for Defense Research and Development (DDR\&D), the US-Israel Binational Science Foundation (BSF) and US National Science Foundation (NSF), the Minerva Foundation with funding from the Federal German Ministry for Education and Research, the Center for New Scientists at the Weizmann Institute of Science, the Council for Higher Education (Israel), QUANTERA (PACE-IN), the Helmsley Charitable Trust, and the Estate of Louise Yasgour. This research is made possible in part by the historic generosity of the Harold Perlman Family.
\end{acknowledgments}




\end{document}